\documentclass[floatfix,superscriptaddress,amsmath,amssymb,notitlepage,twocolumn,aps]{revtex4-1}

\newcommand{\vect}[1]{\boldsymbol{#1}}

\usepackage{cmbright}
\DeclareFontShape{OT1}{cmss}{m}{it}{<->ssub*cmss/m/sl}{}

\usepackage{textcomp,gensymb}
\usepackage{lmodern}
\usepackage{amsmath}
\usepackage{amssymb}

\usepackage[noindentafter]{titlesec}
\usepackage{titlesec}
\titleformat{name=\section}
  {\normalfont\large\bfseries\MakeUppercase}{\MakeUppercase{\thesection}}{0pt}{}
\titleformat{name=\subsection}
  {\normalfont\bfseries}{\thesection}{0pt}{}
\titlespacing{\section}{0cm}{0.7cm}{0.01cm}
\titlespacing{\subsection}{0cm}{0.45cm}{0cm}

\usepackage[utf8]{inputenc}
\usepackage{listings}
\usepackage{footmisc}
\usepackage{enumerate}
\usepackage{latexsym}
\usepackage{braket}
\usepackage[version=4]{mhchem}
\usepackage{graphicx}
\usepackage[caption=false]{subfig}
\usepackage[colorlinks=True,linkcolor=red,citecolor=blue,urlcolor=blue]{hyperref}
\usepackage{blkarray}
\usepackage{array}
\usepackage[dvipsnames]{xcolor}
\usepackage[normalem]{ulem}
\usepackage{wasysym}
\usepackage{multirow}
\usepackage{siunitx}

\usepackage{xspace}
\usepackage{subfiles}

\usepackage{tabularx}
\usepackage{array}   
\newcolumntype{L}{>{$}l<{$}} 
\newcolumntype{R}{>{$}r<{$}} 
\newcolumntype{C}{>{$}c<{$}} 

\newcommand{\beginsupplement}{%
        \setcounter{table}{0}
        \renewcommand{\thetable}{S\arabic{table}}%
        \setcounter{figure}{0}
        \renewcommand{\thefigure}{S\arabic{figure}}%
     }
\usepackage{float}

\usepackage[capitalise]{cleveref}
\usepackage{placeins}

\date{\today}

\hypersetup{
    colorlinks=true,
    linkcolor=blue,
    filecolor=magenta,      
    urlcolor=blue,
    pdftitle={Overleaf Example},
    pdfpagemode=FullScreen,
    }

\usepackage{comment}

\usepackage{float}

\usepackage{epstopdf}
\AppendGraphicsExtensions{.tif}

\begin{document}
\title{Absence of magnetic order in \texorpdfstring{\ce{RuO_2}}{RuO2}: insights from \texorpdfstring{$\mu$SR}{muSR} spectroscopy and neutron diffraction}

\author{Philipp Keßler}
\email{philipp.kessler@uni-wuerzburg.de}
\altaffiliation[]{\newline These two authors contributed equally.}
\affiliation{Physikalisches Institut, Universität Würzburg, 97074 Würzburg, Germany}
\affiliation{Würzburg-Dresden Cluster of Excellence ct.qmat, Universität Würzburg, 97074 Würzburg, Germany}

\author{Laura Garcia-Gassull}
\email{gassull@itp.uni-frankfurt.de}
\altaffiliation[]{\newline These two authors contributed equally.}
\affiliation{Institut f\"ur Theoretische Physik, Goethe-Universit\"at, 60438 Frankfurt am Main, Germany}

\author{Andreas Suter}
\author{Thomas Prokscha}
\author{Zaher Salman}
\email{zaher.salman@psi.ch}
\affiliation{Laboratory for Muon Spin Spectroscopy, Paul Scherrer Institute, CH-5232 Villigen PSI, Switzerland}

\author{Dmitry Khalyavin}
\author{Pascal Manuel}
\email{pascal.manuel@stfc.ac.uk}
\author{Fabio Orlandi}
\affiliation{ISIS Pulsed Neutron and Muon Source, STFC Rutherford Appleton Laboratory, Didcot, OX11 0QX, United Kingdom}

\author{Igor I. Mazin}
\email{imazin2@gmu.edu}
\affiliation{George Mason University, Department of Physics \& Astronomy and  Quantum Science and Engineering Center,  Fairfax, USA}

\author{Roser Valent\'i}
\email{valenti@itp.uni-frankfurt.de}
\affiliation{Institut f\"ur Theoretische Physik, Goethe-Universit\"at, 60438 Frankfurt am Main, Germany}

\author{Simon Moser}
\email{simon.moser@uni-wuerzburg.de}
\affiliation{Physikalisches Institut, Universität Würzburg, 97074 Würzburg, Germany}
\affiliation{Würzburg-Dresden Cluster of Excellence ct.qmat, Universität Würzburg, 97074 Würzburg, Germany}

\date{\today}

\begin{abstract}
Altermagnets are a novel class of magnetic materials besides ferro- and antiferromagnets, where the interplay of lattice and spin symmetries produces a magnetic order that is staggered both in coordinate as well as momentum space. The metallic rutile oxide \ce{RuO2}, long believed to be a textbook Pauli paramagnet, recently emerged as a workhorse altermagnet when resonant X-ray and neutron scattering studies reported nonzero magnetic moments and long-range collinear order. While experiments on thin films seem consistent with altermagnetic behavior, the origin and size of magnetic moments in \ce{RuO2} still remain controversial. Here we show that \ce{RuO2} is nonmagnetic, regardless if as bulk or thin film. Employing muon spin spectroscopy as a highly sensitive probe of local magnetic moments complemented by density functional theory, we find at most \SI{1.4e-4}{\micro_B}/\ce{Ru} in bulk \ce{RuO2} and at most \SI{7.5e-4}{\micro_B}/\ce{Ru} in epitaxial films. In their essence, these moments reflect the detection limit of our spectrometers and are orders of magnitude smaller than previously
reported neutron results, i.e., the moments previously assumed to rationalize altermagnetic behavior. Our own neutron diffraction measurements on \ce{RuO2} single crystals identify multiple scattering as a likely source for this discrepancy. 
\end{abstract}

\maketitle
\section{Introduction}

Altermagnetism, a magnetic state of matter that shares properties with both conventional ferro- and antiferromagnets \cite{editorial,PRXreview}, has recently garnered considerable attention. While altermagnets exhibit the typical ferromagnetic signature of spin split energy bands, they still retain the zero net magnetization that is characteristic of an antiferromagnet \cite{Cuono}. Since one of the first articles on the subject \cite{LiborRuO2}, a number of theoretical and experimental studies aimed at exploring fingerprints of altermagnetism have been conducted \cite{Kartik2020, feng_anomalous_2022, Betancourt2023, Tschirner2023a, Liu2024, Reimers2024, Zhou2024, Hariki2024, Guo2024}, 
but a definitive reproducible experimental detection of the altermagnetic state remains elusive. A plurality of experiments, in particular, have been performed on the metallic rutile oxide \ce{RuO2} \cite{Occhialini2021, Liang,Karube2022,feng_anomalous_2022, Tschirner2023a, Fedchenko2024, Hiraishi2024, Prateek2024, Guo2024}, where both resonant X-ray scattering \cite{Zhu} and neutron diffraction experiments \cite{berlijn_itinerant_2017} reported a $\vect{q} = \vect{0}$ antiferromagnetic pattern that is consistent with altermagnetic symmetry constraints. Nonetheless, in the earlier neutron report \cite{berlijn_itinerant_2017}, two caveats were noted: firstly, polarized neutron diffraction refinement resulted in an ordered moment of approximately \SI{\sim 0.05}{\micro_B}/Ru. This value is at least one order of magnitude smaller than the numbers used to interpret subsequent experiments \cite{feng_anomalous_2022, Tschirner2023a}, but still two orders of magnitude larger than what was found recently by muon spin relaxation/rotation ($\mu$SR) technique on \ce{RuO2} single crystals \cite{Hiraishi2024}.

Secondly, electronic structure calculations in the framework of density functional theory (DFT) converge, without additional nudging, to a nonmagnetic solution \cite{smolyanyuk2023ruo2, berlijn_itinerant_2017,Liang}.
Only the addition of a sizable Hubbard $U$ -- somewhat uncommon for good 4d metals -- can generate ordered magnetic moments. In this context, it has been suggested that the presence of \ce{Ru} vacancies may lead to magnetism at a lower and more realistic Hubbard $U$ \cite{smolyanyuk2023ruo2} than that used in Refs. \cite{LiborRuO2,ahn2019}.

In view of the popularity and conceptual importance of \ce{RuO2} as a prototypical altermagnet, it is essential to verify the existence and strength of the magnetic order in this compound. We thus have revisited the question about ordered moments in \ce{RuO2} and performed careful neutron diffraction experiments employing state-of-the art time-of-flight detection. Surprisingly, we found that the previously reported and alleged magnetic (1,0,0) reflection \cite{berlijn_itinerant_2017} is likely related to a multiple scattering artifact rather than to a magnetic origin. Further, we employed the highly sensitive $\mu$SR technique, complemented by DFT calculations of the muon stopping sites and hyperfine fields, and determined an upper bound on magnetic moments of \SI{1.4e-4}{\micro_B}/Ru in the bulk and \SI{7.5e-4}{\micro_B} in epitaxial thin films of \ce{RuO2}, consistent with a recent bulk value obtained by an independent group \cite{Hiraishi2024}. 
    
These moments are two orders of magnitude smaller than what was reported by the previous neutron study \cite{berlijn_itinerant_2017}, and essentially reflect the detection limit of our spectrometer. Hyperfine field calculations in stoichiometric and \ce{Ru} vacant \ce{RuO2} further exclude any fortuitous cancellation of hyperfine fields at the muon stopping site. We thus conclude that \ce{RuO2} exhibits no long-range magnetic order, neither in the bulk  nor in epitaxial thin films, and the altermagnetic signatures reported in multiple earlier publications \cite{feng_anomalous_2022, Karube2022, Tschirner2023a, Fedchenko2024,Guo2024,Zhou2024} likely have extrinsic origin.

\section{Experimental Results}
\label{Sec:Results}
\subsection{\texorpdfstring{\ce{RuO2}}{RuO2} Samples}
To shed light on the magnetic properties of \ce{RuO2}, we investigated three different types of samples: (i) Millimeter sized \ce{RuO2} crystals that were grown by chemical vapor transport at the Crystal Growth Facility of the EPFL in Lausanne, Switzerland, following a recipe described in Ref.~\cite{Schafer1963}; (ii) 11 nm thick \ce{RuO2} epitaxial thin films grown on \ce{TiO2}(110) by pulsed laser deposition at the Institute of Physics, Würzburg, Germany, following our recipe developed in Ref.~\cite{Tschirner2023a}; and (iii) \SI{99.9}{\%} pure (trace metals basis) \ce{RuO2} powder, purchased from Sigma-Aldrich that was measured without further preparation steps. The \ce{RuO2} crystalline structure was ensured for all three sample types by x-ray diffraction techniques before the $\mu$SR and/or neutron diffraction experiments.

\subsection{Neutron diffraction}

Reassessing the results of the earlier study \cite{berlijn_itinerant_2017}, we first performed single crystal neutron diffraction on the \ce{RuO2} crystals. Hereby, we employed the large pixelated detector array of the state-of-the-art time of flight diffractometer WISH at the ISIS Neutron and Muon Source of the STFC Rutherford Appleton Laboratory \cite{Chapon2011}. Measurements at \SI{1.5}{K} on four \ce{RuO2} crystals, three of them identified as multi grain, one of them identified as single grain, all led to consistent conclusions. The crystals were aligned in the $(h,0,l)$ horizontal scattering plane, a large portion of which can be measured in a single shot using the time-of-flight dimension, with the possibility to rotate the crystals about the vertical $b^*$ direction. 

\begin{figure}[htb]
\includegraphics[width=0.95\linewidth]{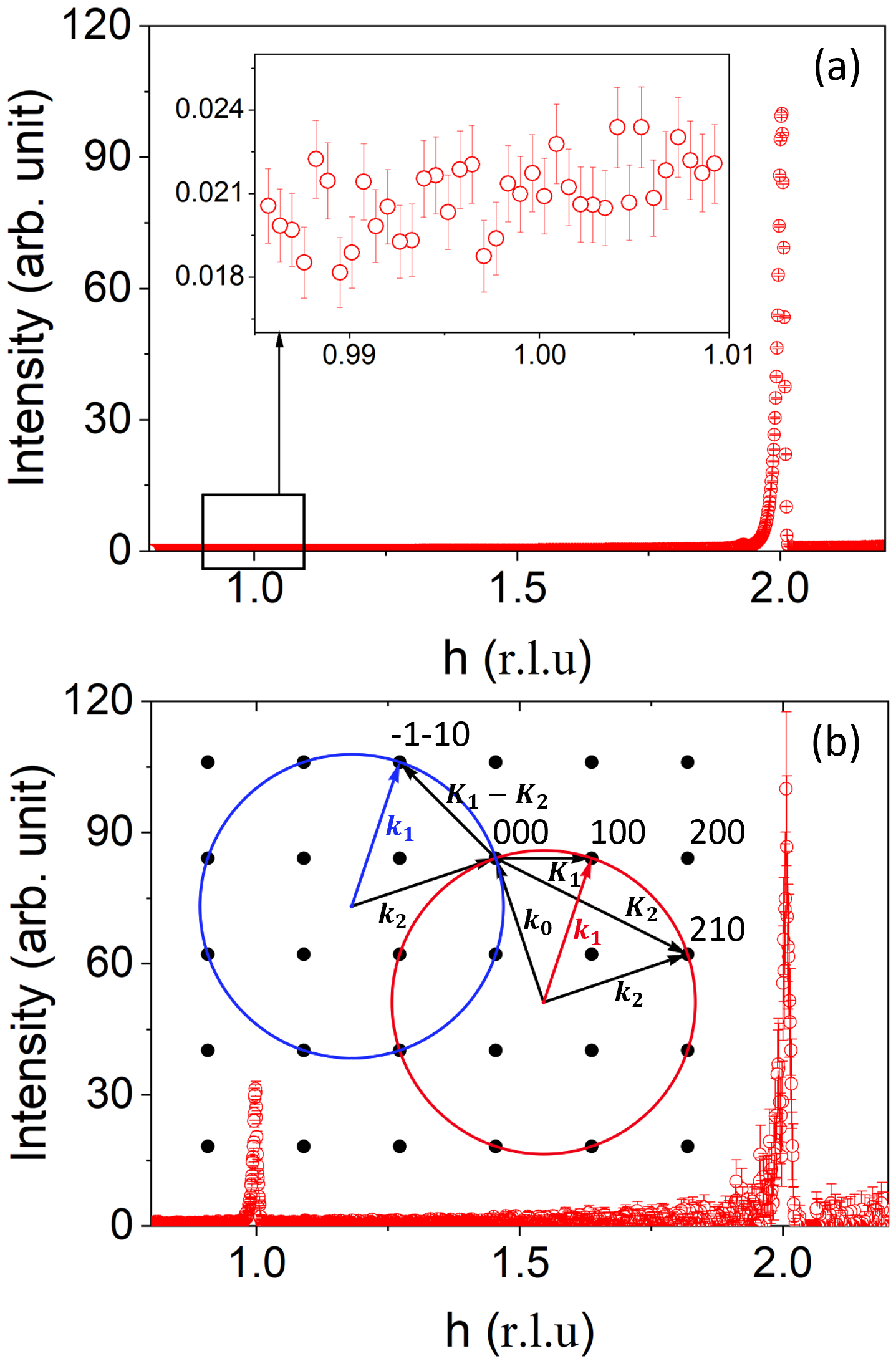}
\caption{Neutron diffraction data collected on \ce{RuO2} single crystal at $T = \SI{1.5}{K}$ with the $(h,0,0)$ reflections at the scattering angle $2 \Theta = \SI{71}{\degree}$ (a) and $2 \Theta = \SI{32}{\degree}$ (b). The counting time was 10 hours and 15 minutes, respectively. The data demonstrate 1D cuts of the reciprocal lattice along the $(h,0,0)$ direction. The inset of panel (a) shows a zoomed region at a vicinity of the $(1,0,0)$ reflection. The inset of panel (b) shows a simple two-dimensional illustration of the double diffraction, taking place when the Ewald sphere intersects both the $\vect{K}_1 = (1,0,0)$ and the $\vect{K}_2=(2,1,0)$ reciprocal lattice points. The beam, first scattered by the $(2,1,0)$ planes in the direction $\vect{k}_2$, is then scattered again by the $(-1,-1,0)$ planes in the direction $\vect{k}_1$, where the $(1,0,0)$ scattering is expected. The intersection of the $(h,k,0)$ reciprocal plane and the Ewald sphere is shown by the red circle for the first scattering process (from the $(2,1,0)$ planes) and by the blue circle for the second scattering process (by the $(-1,-1,0)$ planes).}
\label{fig:neutron_diffraction}
\end{figure}

As discussed above, altermagnetism in \ce{RuO2} implies a $\vect{q} = \vect{0}$ propagation vector and antiparallel magnetic coupling of the two \ce{Ru} sites in the primitive unit cell, with moments polarized along the crystal c-axis. The structure is expected to result in a strong magnetic scattering to the structurally forbidden $(1,0,0)$ reflection \cite{berlijn_itinerant_2017}. This is due to the fact that the two magnetic sites sum over in the corresponding structure factor and the moments are perpendicular to the scattering vector. In addition, among all possible magnetic reflections with $\vect{q} = \vect{0}$, the $(1,0,0)$ one has the largest $d$-spacing and therefore the largest value of the magnetic form factor. We thus focused our attention at the $(1,0,0)$ reflection and measured it at different scattering angles $2\Theta$, thereby using different neutron wavelengths $\lambda$, and found a strong dependence on the scattering geometry. At some scattering angles, the reflection was not observed even after ten hours of statistics, in spite of a good optimisation with respect of the neutron flux, whereas at some other angles (mainly low angles) it was detectable within a few minutes with a poorer flux optimisation (Fig. \ref{fig:neutron_diffraction}). This behaviour is not typical for a real Bragg peak and is known as Renninger effect \cite{Renninger1937,Speakman1965,Giacovazzo2011}, caused by multiple scattering processes (often referred to as double scattering). 

This can happen when two reciprocal lattice nodes, $\vect{K}_1$ and $\vect{K}_2$, are present simultaneously on the surface of the Ewald sphere. Let us denote the Miller indices of the corresponding families of lattice planes as $(h_1,k_1,l_1)$ and $(h_2,k_2,l_2)$, then the corresponding Laue conditions are: $(\vect{k}_1-\vect{k}_0)/\lambda = \vect{K}_1$ and $(\vect{k}_2-\vect{k}_0)/\lambda = \vect{K}_2$, where $\vect{k}_0$ is the unit vector along the incident beam and $\vect{k}_1$, $\vect{k}_2$ are directions of the diffracted beam. By subtracting the second equation from the first, one obtains $(\vect{k}_1-\vect{k}_2)/\lambda=\vect{K}_1-\vect{K}_2$. This indicates that the beam diffracted by the $(h_1,k_1,l_1)$ in the direction $\vect{k}_1$ overlaps with a double-scattering beam, first by the plane $(h_2,k_2,l_2)$ in the direction $\vect{k}_2$ and then by the plane $(h_1-h_2, k_1-k_2, l_1-l_2)$ in the direction $\vect{k}_1$ (see inset of Fig. \ref{fig:neutron_diffraction}(b) for an illustration). If $(h_1,k_1,l_1)$ is a systematically absent structural reflection, then the Renninger effect may result in an apparent violation of the reflection conditions \cite{Sazonov2016} or can be interpreted as magnetic scattering.

As the radius of the Ewald sphere is defined by the wavelength $\lambda$ of the incident and scattered neutrons, the double scattering process is expected to be strongly $\lambda$-dependent and should not take place if the necessary conditions are not satisfied. Observation of the doubly scattered beam will thus only exist in a finite angular range due to beam divergence and finite extension of the Bragg spots in reciprocal space, controlled by sample quality and instrument resolution. At low scattering angles, where the Bragg law selects short wavelengths, the radius of the Ewald sphere is big, and the sphere has a large surface in reciprocal space. Further, instrumental resolution also quickly degrades with decreasing scattering angles. This strongly increases the probability that two (or more) reciprocal spots hit the sphere producing the multiple scattering. Consequently, the Renninger effect is the most natural explanation of the presence of the $(1,0,0)$ reflection in \ce{RuO2} and its strong dependence on the scattering angle (or equivalently the wavelength of neutrons). In the context of the present study, the key experimental observation is the lack of the $(1,0,0)$ reflection in the well-optimised (in respect of the neutron flux) scattering geometry. After ten hours of measurements, the statistic of the data allows us to rule out the $\vect{q} = \vect{0}$ magnetic ordering with moment size bigger than \SI{0.011}{\micro_B}, as previously suggested in Ref.\;\cite{berlijn_itinerant_2017}. Moreover, we explored a large portion of the reciprocal space available in the time-of-flight scattering experiment and have not found any evidence of magnetic ordering either with $\vect{q} = \vect{0}$ or with some other propagation vector. This strongly supports the non-magnetic nature of the \ce{RuO2} crystals.

\subsection{Muon spin resonance \texorpdfstring{($\mu$SR}{muSR})}

For further insights on their size, we employed the $\mu$SR technique as a highly sensitive probe of static local magnetic moments. The $\mu$SR experiments on the powder and single crystalline samples were conducted at the General Purpose Surface-Muon (GPS) Instrument \cite{Amato2017RSI} at the Swiss Muon Source piM3.2 beamline of the Paul Scherrer Institute (PSI). A $^4$He continuous-flow cryostat was used to perform $\mu$SR measurements between \SI{5}{K} and \SI{290}{K}. The thin films were studied in the Low-Energy Muons Facility (LEM) at the $\mu$E4 beamline at PSI \cite{Morenzoni1994PRL,Prokscha2008NIPA}. For these measurements, the thin-film samples were glued onto the cold finger of a $^4$He flow cryostat using silver paint. The data analysis for all the $\mu$SR spectra was conducted using the {\it musrfit} package \cite{Suter2012}.

In a $\mu$SR measurement, fully polarized muons are implanted into the sample. Muons decay with a lifetime of \SI{2.2}{\micro s}, emitting a positron preferentially in the direction of the spin polarization at the time of decay. The spatial distribution of these positrons is monitored as a function of time using plastic scintillating detectors placed around the studied sample. Therefore, the asymmetry in the decay of positrons on opposite sides of the sample, $A(t)$, is proportional to the polarization along the corresponding axis. In practice, $A(t)$ is calculated via
\begin{equation} \label{eq:asymmetry}
    A(t) = \frac{N_\text{U}- \alpha_\text{UD} N_\text{D}}{N_\text{U}+ \alpha_\text{UD} N_\text{D}}, 
\end{equation}
where $N_\text{U}$ and $N_\text{D}$ are the number of positrons detected as a function of time in detectors placed above (U) and below (D) the sample, respectively. $\alpha_\text{UD}$ is a geometric correction factor to account for the different efficiencies of the U and D detectors.

\begin{figure}[htb]
\includegraphics[width=1\linewidth]{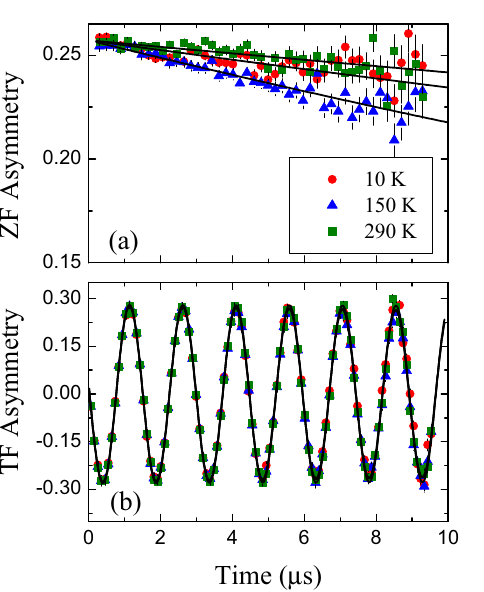}
\caption{The asymmetry measured in (a) zero and (b) transverse magnetic field at various temperatures in bulk \ce{RuO2}.}
\label{fig:Asymmetry}
\end{figure}

In the bulk $\mu$SR measurements (GPS), we performed primarily measurements on the powder sample in zero applied external field (ZF) and in weak transverse field (TF - transverse to the initial muon spin polarization). A few measurements obtained from the single crystal gave results identical to the powder measurements. Therefore, we hereafter focus only on the powder measurements. In Fig.~\ref{fig:Asymmetry}(a), we show typical asymmetry curves measured at three different temperatures.
We observe no significant relaxation or depolarization at any temperature and no missing asymmetry, confirming that the local magnetic fields sensed by the implanted muons are extremely small. Only a very small increase in the relaxation rate around $\sim150$~K is observed. Similar results were obtained from $\mu$SR measurements while applying a TF of \SI{5}{mT}. A few representative asymmetry spectra are depicted in Fig.~\ref{fig:Asymmetry}(b). The applied field in this case leads to a Lamor precession of the muons' spin polarization, and therefore, an oscillation in the asymmetry spectra. The initial amplitude of these oscillations is proportional to the paramagnetic volume fraction of the sample. 
Just as in the case of the ZF measurements, no significant temperature dependence was observed ruling out any static magnetic order in \ce{RuO2} (neither short nor long range). 

\begin{figure}[t!]
\includegraphics[width=1\linewidth]{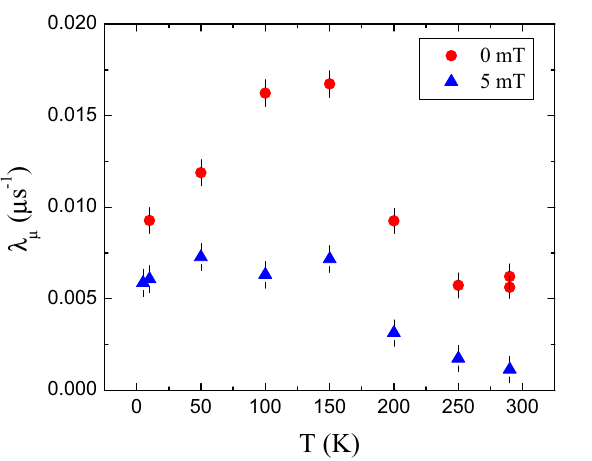}
\caption{Temperature dependence of the muon relaxation rate in bulk \ce{RuO2} for zero-field and transverse-field geometry.}
\label{fig:depolarization}
\end{figure}

To analyse the data, we fit the asymmetries using an exponentially damped oscillation, $A (t) = A_0  \exp(-\lambda_{\mu} t) \cos( \omega t + \varphi)$. Both ZF and TF data were fitted using a temperature independent initial asymmetry $A_0$ and phase $\varphi$, while the relaxation rate $\lambda_{\mu}$ and frequency $\omega$ were temperature dependent. For the ZF measurements, we use $\omega = 0$ and $\varphi = 0$. The relaxation rates obtained from these fits are shown in Fig. \ref{fig:depolarization}. 
We note a very small temperature dependence in $\lambda_{\mu}$ measured in both ZF and TF. In particular, in the ZF measurements we observe a peak around $\sim 150$~K, which corresponds to the small enhancement in relaxation seen in the raw spectra of Fig.~\ref{fig:Asymmetry}(a). In contrast, in the TF measurements, $\lambda_{\mu}$ increases slightly below $\sim 200$~K but saturates below $\sim 150$~K. This small temperature dependence and extremely small relaxation rate, however, cannot be attributed to static magnetic order in \ce{RuO2}, which would generally result in much larger values of $\lambda_{\mu}$.

\begin{figure}[htb]
\includegraphics[width=1\linewidth]{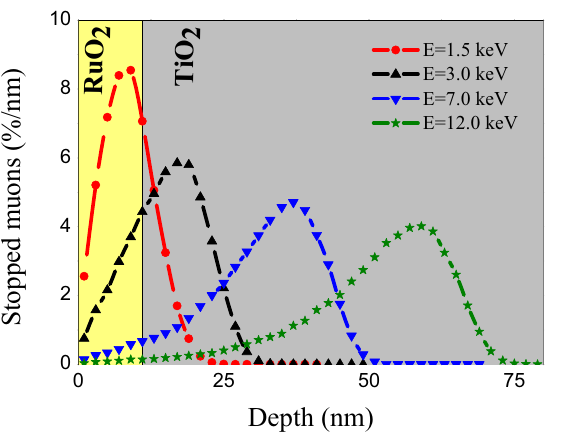}
\caption{Calculated muon stopping profile as a function of depth for the different implantation energies in an 11~nm thick \ce{RuO2} on \ce{TiO2} substrate.}
\label{fig:TrimSP}
\end{figure}

\begin{figure}[htb]
\includegraphics[width=1\linewidth]{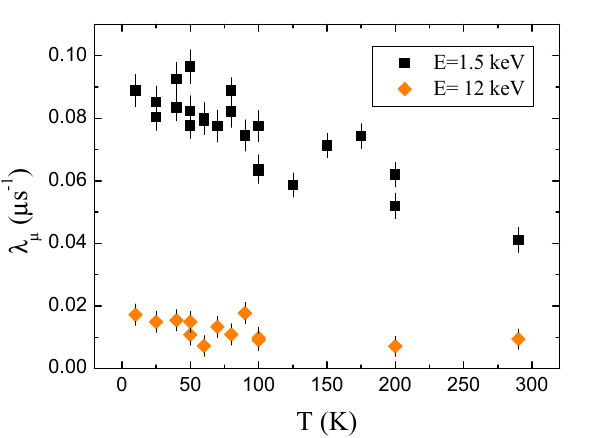}
\caption{Temperature dependence of the muon relaxation rate in a film of 11~nm \ce{RuO2} on \ce{TiO2} measured in 10~mT transverse-field. At $E=1.5$~keV most of the muons stop in the \ce{RuO2} layer while at $E=12$~keV they stop in the substrate.}
\label{fig:LEM}
\end{figure}

In order to probe a possible magnetic order (or static magnetic moments) near grain boundaries, interface, or other defects, we performed low energy $\mu$SR (LE-$\mu$SR) measurements on our 11~nm thick epitaxial \ce{RuO2}(110) thin films deposited on top of a \ce{TiO2} rutile substrate. In the LE-$\mu$SR, the implantation energy of the muons was varied between 1~keV and 12~keV, resulting in an implantation depth (probing depth) between $\sim 5$~nm and $\sim 100$~nm (see Fig.~\ref{fig:TrimSP}).
We performed LE-$\mu$SR measurements in TF of 10~mT as a function of temperature at two implantation energies: $E=1.5$~keV, where most of the muons stop in the \ce{RuO2} layer; and $E=12$~keV, where practically all the muons stop in the \ce{TiO2} substrate. These energies were selected based on TRIM.SP calculations \cite{Morenzoni2002NIMPRSB,Eckstein1991} (Fig.~\ref{fig:TrimSP}). Like the bulk data, the obtained spectra were fitted to an exponentially damped oscillation. The resulting damping rate, $\lambda_{\mu}$, is plotted as a function of temperature for both implantation energies in Fig.~\ref{fig:LEM}.

We find that $\lambda_{\mu}$ in the \ce{RuO2} layer, though generally higher than what we found in the bulk, remains quite small. This rules out the presence of large magnetic fields in these films. We also note that $\lambda_{\mu}$ in \ce{RuO2} is much larger than that measured in the \ce{TiO2} substrate. Therefore, the enhanced relaxation rate in the \ce{RuO2} layer cannot be attributed to the fraction of muons stopping in the substrate. Moreover, the gradual increase in $\lambda_{\mu}$ with decreasing temperature rules out background contribution due to back-scattered muons \cite{Suter2023JPCS}. Instead, we believe that the observed increase in relaxation compared to the bulk is due to the presence of somewhat larger magnetic moments in films of \ce{RuO2}. The origin of these magnetic moments may be defects or vacancies, which might be more abundant in the films.

\section{Discussion}
In order to better interpret the results of the $\mu$SR measurements, we performed first-principles calculations based on spin-polarized density functional theory (DFT), and enforcing a magnetic moment on the \ce{Ru} atoms by adding a large effective Hubbard $U$~\cite{smolyanyuk2023ruo2}, as also done in other computational studies~\cite{LiborRuO2,ahn2019}. To emulate the muon stopping site, we used the standard protocol described in Ref.~\cite{Blundell2022}. As muon and proton are identical from the point of view of adiabatic DFT calculations, we used a proton in place of the muon, thinking of the muon as a light proton \cite{Blundell1999}. The total number of electrons was kept unchanged. This proton was implanted 
into a sufficiently large ($3\times3\times3$) supercell to avoid interaction with its replicas in the neighbouring cells \cite{Blundell2022}. 
Otherwise, we performed DFT calculations with the projector augmented-wave (PAW) method~\cite{Kresse1999, Blochl1994}, as implemented in \textit{VASP} \cite{kresse_ab_1993,kresse_ab_1994,kresse_efficiency_1996,kresse_efficient_1996,perdew_generalized_1996,perdew_generalized_1997}. As exchange-correlation functional we considered the generalized gradient approximation (GGA)~\cite{Ernzerhof} and added an effective Hubbard $U_{eff}=U-J=1.4$~eV using the rotational invariant GGA+U introduced in Ref. \cite{Liechtenstein} to ensure a sizable magnetic moment on the \ce{Ru} atoms and a non-zero hyperfine field. The cutoff energy for the plane-wave basis was set to 300~eV.

\begin{figure}[t!]
\includegraphics[width=0.9\linewidth]{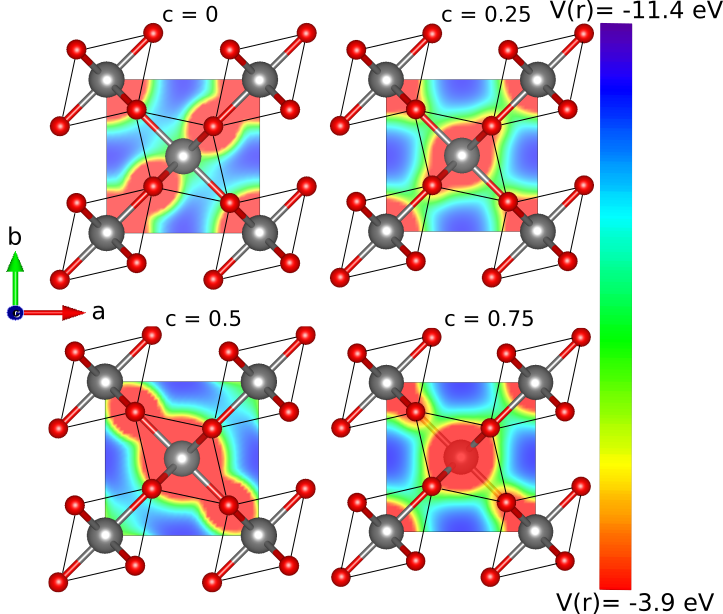}
\caption{Crystal structure and electrostatic potential maps for different cuts along c in units of the lattice parameter. The electrostatic potential is represented by the color map from red to blue. The purple shade indicates regions where the (negative) electrostatic potential is the largest, i.e., the likely muon stopping sites. The color map is limited to the unit cell.}
\label{fig:evolution_locpot_along_z}
\end{figure}

\ce{RuO2} crystallizes in the rutile structure, space group P4$_2$/mnm, as depicted in Fig. \ref{fig:crystal_structure}. This structure, which we here call ``pristine'', serves as the basis for DFT calculations involving a muon. The calculated magnetic moment on the \ce{Ru} atoms, for $U_{eff}=1.4$~eV, was 0.866 $\mu_B$/Ru. This magnetic moment was used in the interpretation of the experiments done in Refs.~\cite{berlijn_itinerant_2017, Zhu, Tschirner2023a, Kartik2020, Adamantopoulos2024, feng_anomalous_2022, Zhou2024, Guo2024}. In addition to the pristine structure, we emulated  the effect of potential \ce{Ru} vacancies by removing one \ce{Ru} from this supercell (i.e., 1.8~\% of \ce{Ru} vacancies as reported in at least one case \cite{vacancies_found}), which enhances the tendency to magnetism \cite{smolyanyuk2023ruo2}. 
We also checked the stability of the results against adding spin-orbit coupling (SOC), and found that both in the pristine case and in the presence of a \ce{Ru} vacancy, SOC introduces negligible effects, in agreement with the previous literature \cite{Jovic2018, ahn2019}. To optimize computing time we thus performed our calculations without SOC.

First, we identified probable stopping sites for a muon by looking at the electrostatic potential in the pristine unit cell, as shown in Fig. \ref{fig:evolution_locpot_along_z}. This electrostatic potential is defined as the sum of the ionic potential and the Hartree potential.

\begin{figure}[htb]
\includegraphics[width=1\linewidth]{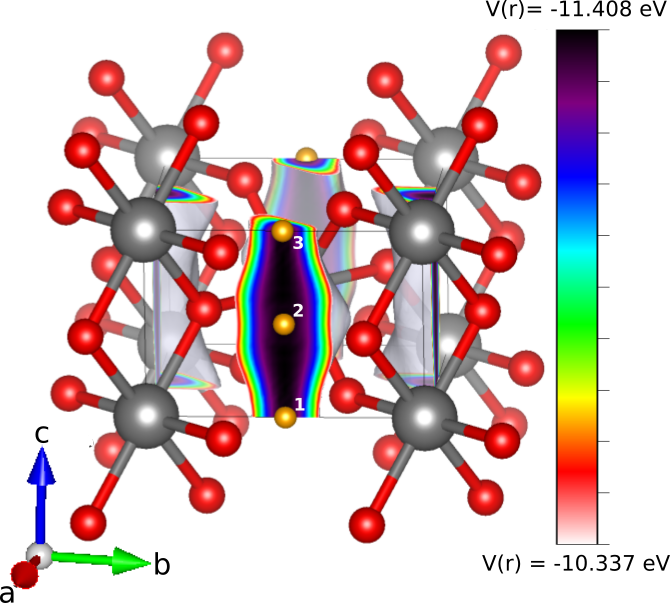}
\caption{Initial approximation of the stopping sites of the muon, represented by orange balls, where the electrostatic potential is highest. The color map indicates that, the maxima of the potential occurs at the furthest distance from \ce{Ru} atoms. These positions are the most likely stopping sites along the channel, as seen in the minima of the energy in Fig. \ref{fig:diffusion}.
 }
\label{fig:Prob_stopping_sites}
\end{figure}

As Fig.~\ref{fig:evolution_locpot_along_z} shows, and Fig.~\ref{fig:Prob_stopping_sites} further highlights, the electrostatic potential suggests the presence of channels along the $c$ direction,  where we expect the muon to be initially attracted to. Within a channel, the electrostatic potential minima are the points farthest away from the \ce{Ru} ions. They are indicated in Fig.~\ref{fig:Prob_stopping_sites} by little orange balls. 

\begin{figure}[t!]
\includegraphics[width=0.99\linewidth]{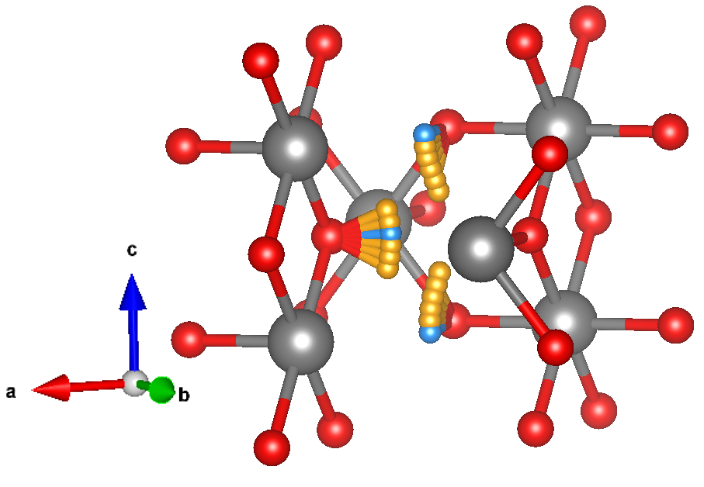}
\caption{Final stopping sites of the muon (represented in orange or blue) for different positions along the $c$ axis. In blue we specify those positions with the lowest energy, i.e. the most likely stopping sites.}
\label{fig:Final_stopping_sites}
\end{figure}

\begin{figure}[t!]
\includegraphics[width=0.99\linewidth]{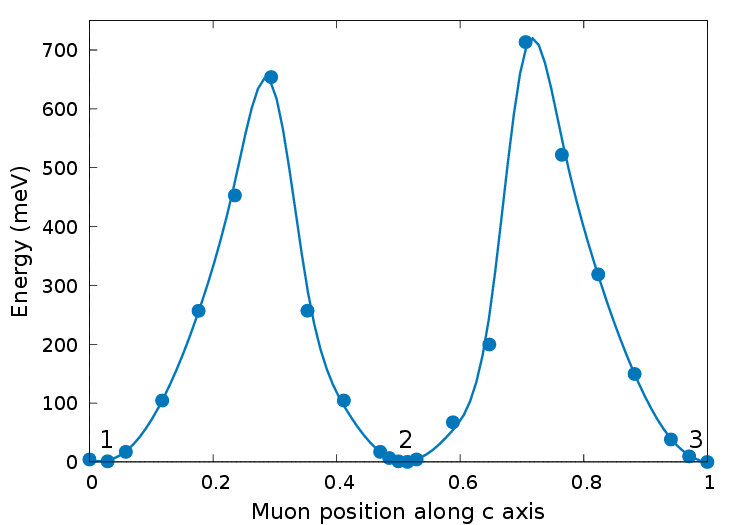}
\caption{Energy calculated for a diffusion path of a muon along the $c$ direction from one \ce{Ru} spin-equivalent plane to the next, in fractional coordinates with \(U_{eff} = 1.4\) eV. The muon's position along $a$ and $b$ was relaxed. The most likely stopping sites are in the planes of the \ce{Ru} atoms, as labeled in Fig \protect\ref{fig:Prob_stopping_sites}. The line is a guide to the eye. Please note that the energy landscape should be periodic, and the observed deviations on the order of $\pm 50$ meV serve as an independent gauge of reliability of our calculations.}
\label{fig:diffusion}
\end{figure}

Taking a closer look at Fig.~\ref{fig:Prob_stopping_sites}, and focusing on one color, i.e.,  one value of the electrostatic potential, we can construct an equipotential surface that is oriented along the $c$ axis (see Fig. \ref{fig:crystal_structure} for an example). The tube-like shape indicates the above-mentioned channel to which a muon will be immediately attracted. Full optimization of the muon, \ce{Ru} and \ce{O} positions further shows that the muons displace away from the axis of the tube to form a bond with one of the nearby oxygens (see Fig.~\ref{fig:Final_stopping_sites}), as is typical in ionic crystals \cite{Blundell2022}. 
Although along the c-axis the radii of the tubes diminish in the region between two \ce{Ru} layers, there is a possibility of a thermally-activated muon diffusion along these channels, which can affect the muon depolarization by averaging the local magnetic fields. In order to check this, we performed constrained optimization, fixing the $z$ coordinate of the muon and relaxing all other internal parameters. Since this is a rather time-consuming procedure, we performed all calculations on an $3\times 3\times 4$ $k$-mesh, starting from a low $k$-mesh density and increasing it as the calculations converged. The result is shown in Fig. \ref{fig:diffusion}. We find the minima labeled 1-3 in Fig.~\ref{fig:Prob_stopping_sites} to be at least 600 meV deep. This barrier height ensures that muons do not diffuse between these stopping sites under our experimental conditions. 

Beyond the pristine system, long range static magnetic order in \ce{RuO2} was suggested to be induced by hole doping due to \ce{Ru} vacancies \cite{smolyanyuk2023ruo2}. To account for this possibility, we conducted two sets of calculations: with and without a \ce{Ru} vacancy in the $3\times3\times3$ supercell.
In the former case, we considered both the same stopping site as in a pristine crystal, again with full re-optimization, and a second case where the muon was put near a vacancy, followed by additional atomic relaxation (see Supplementary Information). As seen in Fig. \ref{fig:vac_U_1_4_tubes_top_view_cut_shows_vacancy}, the electrostatic potential increases at the \ce{Ru} vacancy (in absolute values). When placing a muon nearby the vacancy, however, it converged to a position further away from the \ce{Ru} atoms than in the pristine case, revealing no new energetically favorable stopping sites. 

\begin{figure}[t!]
\includegraphics[width=0.99\linewidth]{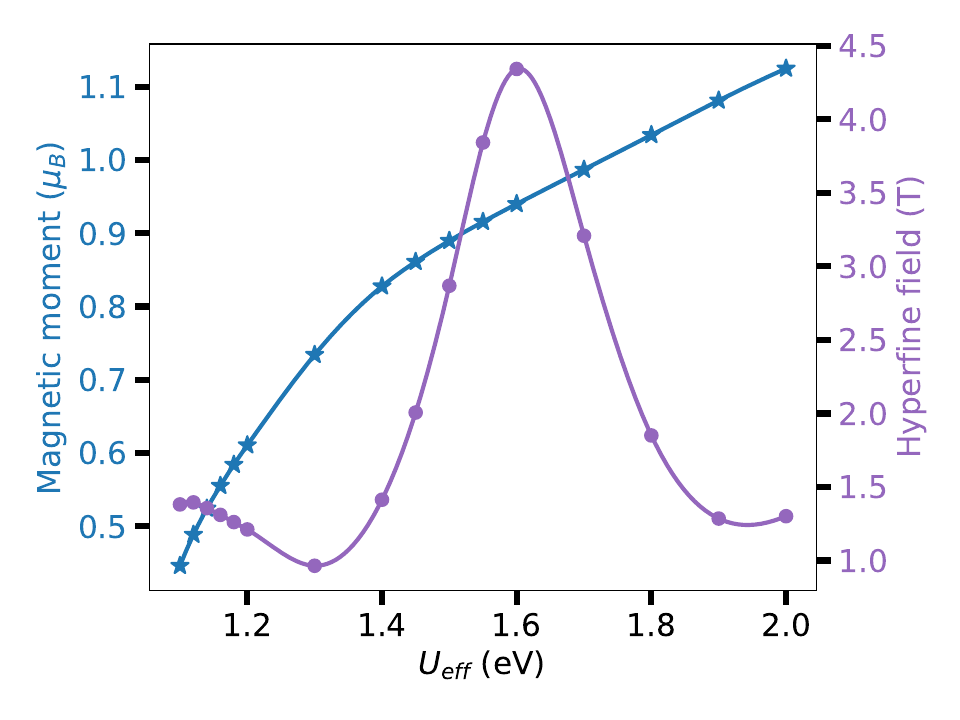}
\caption{Hyperfine field and magnetic moment on \ce{Ru} as a function of Hubbard $U_{eff}$. The Fermi-contact
term provides the dominant contribution to the hyperfine field, being the dipole field contribution negligible. 
For \(U_{eff}\) values smaller than the plotted ones, the \ce{Ru} atoms have no magnetic moment. The lines are guides to the eye.}
\label{fig:hff}
\end{figure}

The hyperfine field acting on the muon was calculated for both the structures with and without \ce{Ru} vacancies. For different $U_{eff}$ in the pristine case, the values of the hyperfine field and the magnetic moment of the \ce{Ru} atoms are shown in Fig. \ref{fig:hff}. As expected in a good metal, we find that the hyperfine field is dominated by the Fermi-contact term due to the finite polarization of the electronic cloud, and the dipole field is much smaller. The hyperfine field shows a non-monotonic dependence on the applied Hubbard $U_{eff}$ and the resulting moment, but never drops below $\approx 1$ T. The calculations with a vacancy did not generate a new stopping site, and the calculated hyperfine field was similar to that for a pristine crystal. That is to say, when the calculations are artificially nudged into a magnetic solution by using a large $U_{eff}$, the resulting Fermi-contact hyperfine field is a few orders of magnitude stronger than admissible by the experimental constraints. Therefore, we conclude that the nonmagnetic calculations without $U$ (as typical for 4d metals) describe our samples better. Note that in the latter case (and in fact for any $U_{eff}\alt 0.6$ eV) the calculations converge to a fully nonmagnetic state, where not only the net magnetic moment around \ce{Ru} is zero, but also the local spin density is zero everywhere.

Given the large Fermi-contact hyperfine field, we can confidently exclude the presence of anti(alter-)magnetic order. However, even if we exclude metallic electrons at the Fermi level from consideration, the dipole field induced by magnetic moments on \ce{Ru} already allows us to exclude any static magnetic order.
Indeed, the knowledge of the muon stopping site can be used to estimate the relationship between the muon relaxation rate $\lambda_{\mu}$ and the magnitude of magnetic moments $m$ in \ce{RuO2}. Due to the small magnitude of $\lambda_{\mu}$ and exponential relaxation of the asymmetry, we assume a dilute arrangement of randomly oriented static moments in our model. Even if the large Fermi-contact field was averaged out, these random moments (located on \ce{Ru} sites or on magnetic impurities) would still produce a dipolar magnetic field distribution that leads to a relaxation/damping of the polarization of implanted muons. 

Considering the exponential relaxation observed experimentally, we assume 
that the muons in \ce{RuO2} experience a distribution of static magnetic fields $f(|B|)$ that can be described by a Lorentzian function, 
\begin{equation}
    f(|B|) = \frac{1}{\pi^2} \frac{\Lambda}{(\Lambda^2 + B^2)^2} 4\pi B^2\;,
\end{equation}
where $\Lambda$ is the half-width at half-maximum. In a physical context, $\Lambda$ can be regarded as the magnitude of the local field $B_\text{loc}$, which can be written in terms of the relaxation rate $\lambda_{\mu}$ in ZF,
\begin{equation}
 B_\text{loc} \approx \Lambda \approx \frac{3}{4} \frac{\lambda_{\mu}}{\gamma_\mu},
\end{equation}
where $\gamma_\mu = 851.615$~MHz/T is the muon's gyromagnetic ratio. Therefore, we can put an upper limit on the local static field experienced by muons of $\SI{1.3e-5}{T}$ in the bulk ($\lambda_{\mu} \sim 0.015$~$\mu$s$^{-1}$) and $\SI{7.0e-5}{T}$ in the films ($\lambda_{\mu} \sim 0.08$~$\mu$s$^{-1}$). Assuming that these fields stem from a dipolar interaction between the muon and a neighbouring \ce{Ru} atom, they can be translated into upper limits on the magnetic moment of $\SI{1.4e-4}{\micro_B}$ in bulk and $\SI{7.5e-4}{\micro_B}$ in films.

\section{Conclusions}

Both neutron scattering experiments on single crystals and $\mu$SR experiments on single crystals, powder and thin films conclusively show that the magnetic moments detected in \ce{RuO2} are several
orders of magnitude smaller than previously suggested by polarized neutron scattering ($\SI{0.05}{\micro_B}$) \cite{berlijn_itinerant_2017}, and essentially comparable to the detection limit of the individual spectrometer facility. In particular,
both the absence of a magnetic Bragg peak in neutron scattering and the measured damping rates of muon spin polarization in our $\mu$SR experiments, present a clear evidence of the non-magnetic ground state of \ce{RuO2}. This is in agreement with the DFT+U results calculated for Hubbard $U$ values that are realistic for the \ce{RuO2} compound \cite{smolyanyuk2023ruo2}.
Our findings strongly suggest that the \ce{RuO2} altermagnetic signatures reported in multiple previous studies likely have extrinsic origin.


\begin{acknowledgments}
The authors thank Helena Reichlova, Dominik Kriegner, Libor Šmejkal, Thomas Jungwirth and Huibo Cao for helpful discussions. Funding support came from the Deutsche Forschungsgemeinschaft (DFG, German Research Foundation) under Germany’s Excellence Strategy through the Würzburg-Dresden Cluster of Excellence on Complexity and Topology in Quantum Matter ct.qmat (EXC 2147, Project ID 390858490) and through the Collaborative Research Center SFB 1170 ToCoTronics (Project ID 258499086). L.G.G. and R.V. further gratefully acknowledge support by the Deutsche Forschungsgemeinschaft (DFG, German Research Foundation) for funding through project TRR 288 — 422213477 (project B05). This work is partially based on experiments performed at the Swiss Muon Source (S$\mu$S), Paul Scherrer Institute, Villigen, Switzerland. I.I.M. was supported by the Army Research
Office under Cooperative Agreement Number W911NF-
22-2-0173. He also acknowledges Heraeus Foundation for supporting his visits to the University of Frankfurt.
\end{acknowledgments}

\section{Author contribution}
P.K. and L.G.G. contributed equally. I.I.M., R.V., S.M. conceived the project. P.K. and S.M. synthesized and characterized the \ce{RuO2} thin films. D.K. and P.M. performed the neutron scattering experiments and analysis. P.K., S.M., Z.S., A.S. and T.P. performed the \(\mu\)SR experiments and analysis and L.G.G. and I.I.M. performed the DFT+U+\(\mu\) calculations. All authors contributed to the discussions and the writing of the paper.

\bibliographystyle{naturemag}

\newpage
\section{Supplementary material}
\begin{figure*}[!h]
\includegraphics[width=0.26\linewidth]{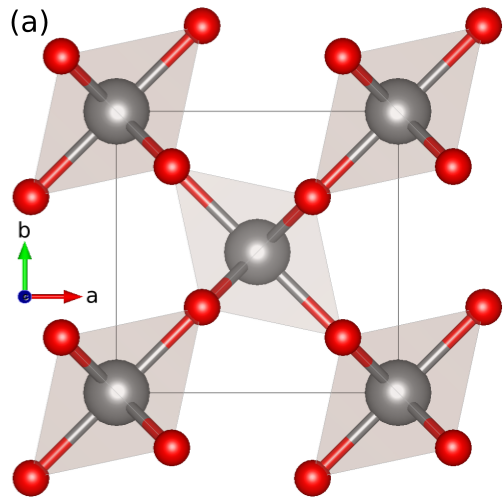}
\includegraphics[width=0.3\linewidth]{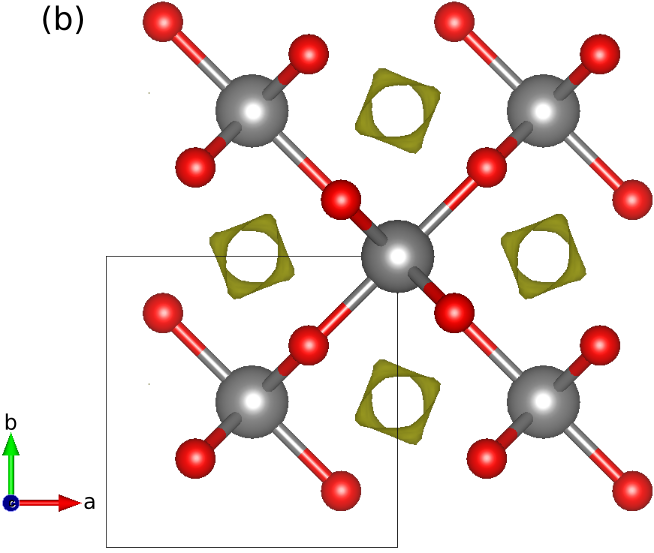}
\includegraphics[width=0.33\linewidth]{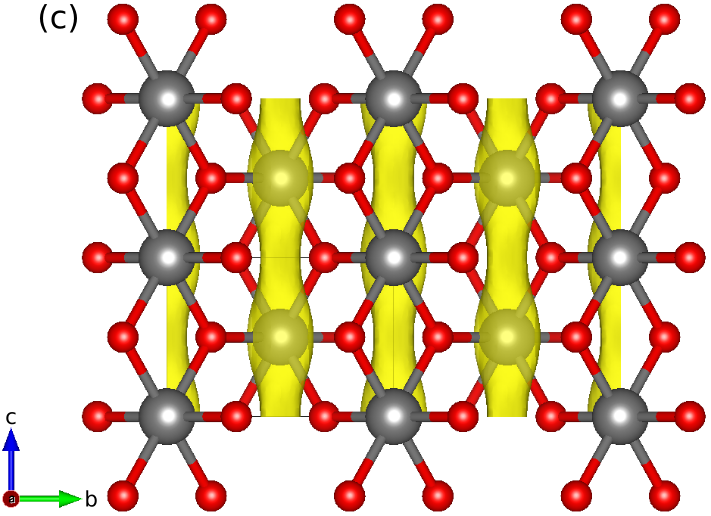}
 \caption{(a): Crystal structure of \ce{RuO2}: \ce{Ru} atoms are represented in gray  while O atoms are in red. The projection on the $ab$ plane of the rotated RuO$_6$ octahedra is shown in shaded gray (note a 90$^\circ$ rotation of the octahedra around  two sublattice Ru). The square shows the unit cell used in calculations without a muon. (b): Top view of the isosurface of the Coulomb potential with value \(-11.2\) eV calculated with $U_{eff}=1.4$ eV, indicating the initial approximation for muon stopping sites. (c): Same, a side view}
\label{fig:crystal_structure}
\end{figure*}

\beginsupplement
\ce{RuO2} crystallizes in the rutile crystal structure, space group P\(4_2/\)mnm. The unit cell of the crystal is shown in the left of Fig. \ref{fig:crystal_structure}, such that the unit cell is formed by two \ce{Ru} atoms, to account for the spin polarized component, and four oxygen atoms. The octahedra formed by the oxygens around the \ce{Ru} atoms are shown in light gray color, exposing the four-fold rotation from one \ce{Ru} sub-lattice to the other. We used this structure to study the behavior of the local potential in the pristine case. The relaxed lattice parameters used are \(a=b= 4.533132\) \r{A} and \(c = 3.124167\) \r{A}.

The middle and right panels of Fig. \ref{fig:crystal_structure} show different perspectives of an isosurface of the electrostatic potential within the unit cell. The top view of the middle panel shows the maxima of the potential to occur in between the \ce{Ru} rows, outside of the octahedra shown in the left panel. Moreover, there is a slight tilt in the isosurface due to the coordination of the nearby oxygens. The right panel shows the tube-like shape of the isosurface along the crystal \(c\)-axis, as well as the supercell used in the calculations. The muon decays somewhere along this tube. The reason to use a supercell is such that the interaction between a muon and its replica in the neighbouring cell is negligible \cite{Blundell2022}. 

\begin{figure}[h]
\includegraphics[width=0.99\linewidth]{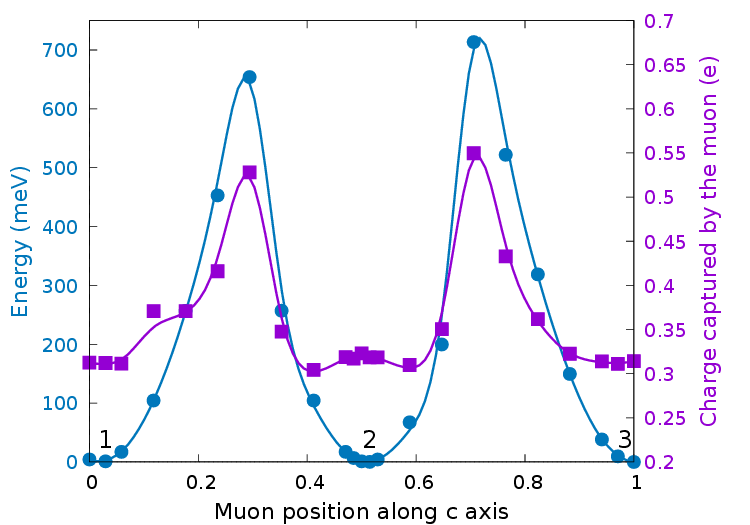}
\caption{Energy of the system (in meV) -left axis- and charge captured by the muon (in electron units) -right axis- for different positions of the muon along the $c$ direction (the positions are given in fractionalized units of the c lattice parameter). The charge the muon is captured from the bond it creates with the nearest O atom, as shown in Fig. \ref{fig:Final_stopping_sites}. The closer to the oxygen atom the muon is, the more screening the muon has and therefore the captured charge is lower, which occurs when the energy of the system is the lowest as well.}
\label{fig:energy_and_charge_vs_muon_z}
\end{figure}

Fig. \ref{fig:energy_and_charge_vs_muon_z} shows the energy of the system, compared to the ground state energy, as well as the charge captured by the muon for different positions along the z axis, in fractionalized coordinates. The minima of the energy occur when closest to the oxygens. Multiple calculations were performed in order to avoid finding a local minimum of the potential energy surface. On the other hand, the muon capturing charge from the environment is a possible occurrence \cite{Blundell2022}. The evolution of the charge captured by the muon shows that, when it is closest to the oxygens (see Fig. \ref{fig:Final_stopping_sites}), the electron screening is higher and hence there is a lower charge captured by the muon.

\begin{figure}[h]
\includegraphics[width=0.99\linewidth]{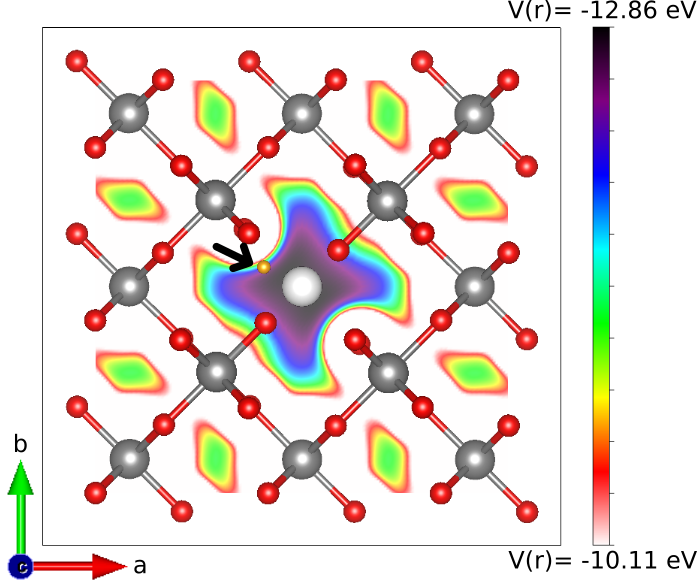}
\caption{Top view of a color map of the electrostatic potential of a pristine crystal with a \ce{Ru} vacancy. Here, the \(U_{eff}\) is \(1.4\) eV. A muon, represented in orange, was placed nearby the $Ru$ vacancy, shown in white, to check for possible new stopping sites. The arrow points to the muon, which is shown in the final stopping site, where it sits near an $O$ atom and creates a bond with it.}
\label{fig:vac_U_1_4_tubes_top_view_cut_shows_vacancy}
\end{figure}

Fig. \ref{fig:vac_U_1_4_tubes_top_view_cut_shows_vacancy} shows a supercell of \(3 \times 3 \times 3\) with one \ce{Ru} vacancy, represented in white centered in the unit cell. This vacancy accounts for \(\approx 2\%\) of vacancies, as reported in Ref. \cite{vacancies_found}. A color map is used to represent the electrostatic potential, with the previously seen absolute maxima in between the octahedra. Moreover, the new electrostatic potential due to the \ce{Ru} vacancy is shown. Note the breaking of \(C_4\) symmetry from the oxygens that would have been bonded to the missing \ce{Ru}. Those oxygens are the ones with the largest displacement, \(0.0953\) \r{A}, towards the vacancy. Furthermore, around the vacancy, the electrostatic potential indicates a possible new stopping site. Hence, a muon was placed nearby the vacancy, more specifically in the (001) plane with the vacancy, to study the energy of the system and compare it to the previously found stopping sites to determine if it was likely the muon ended there. 
The muon was placed in this plane given the knowledge of its preference to bond with the oxygen and the previously minima energy found in the pristine case. Note that the final stopping site of the muon was not at the absolute maxima of the electrostatic potential, due to this tendency to bond with an \(O\) atom. However, no new possible stopping sites that were energetically favorable were found. Therefore, the distance between the most likely stopping sites of the muon found in Fig. \ref{fig:Final_stopping_sites} and the closest \ce{Ru} atom was used to analyze the data of the $\mu$SR experiments.

\end{document}